\documentclass[preprint,sort&compress]{elsarticle}
\usepackage[utf8]{inputenc}
\usepackage[english]{babel}
\usepackage{wrapfig}
\usepackage{amsmath}
\usepackage{amsfonts}
\usepackage{amssymb}
\usepackage{makeidx}
\usepackage{graphicx}
\usepackage{lipsum}
\usepackage{lmodern}
\usepackage{hyperref}
\usepackage{setspace}
\usepackage{fourier}
\usepackage[left=2cm,right=2cm,top=2cm,bottom=2cm]{geometry}\usepackage[utf8]{inputenc}
\usepackage{subcaption}
\usepackage{comment}
\usepackage{color}

\addto\captionsenglish{}

\journal{Journal of Sound and Vibrations}

\let\orgautoref\autoref

\renewcommand{\autoref}[1]
{%
\def\figureautorefname{Fig.}%
\def\subfigureautorefname{Fig.}%
\def\equationautorefname{Eq.}%
\orgautoref{#1}%
}

\begin{document}

\begin{frontmatter}
\title{Theoretical band-gap bounds and coupling sensitivity for a periodic medium with branching resonators}

\author{Mary V. Bastawrous \fnref{fn1}}
\ead{mary.bastawrous@colorado.edu}
\fntext[fn1]{Ann and HJ Smead Department of Aerospace Engineering Sciences}
\author{Mahmoud I. Hussein \fnref{fn1,fn2}}
\ead{mih@colorado.edu}
\fntext[fn2]{Department of Physics}
\address{University of Colorado Boulder, Boulder, CO, 80303}

\begin{abstract}





Elastic metamaterials may exhibit band gaps at wavelengths far exceeding feature sizes. 
This is attributed to local resonances of embedded or branching substructures. 
In branched configurations, such as a pillared plate, waves propagating in the base medium–e.g., the plate portion–experience attenuation at band-gap frequencies. 
Considering a simplified lumped-parameter model for a branched medium, we present a theoretical treatment for a periodic unit cell comprising a base mass-spring chain with a multi-degree-of-freedom, mono-coupled branch. 
Bloch’s theorem is applied, combined with a sub-structuring approach where the resonating branch is modelled separately and condensed into its effective dynamic stiffness. 
Thus, the treatment is generally applicable to an arbitrary branch regardless of its size and properties. 
We provide an analysis$-$supported by guiding graphical illustrations$-$that yields an identification of fundamental bounds for the band-gap edges as dictated by the dynamical characteristics of the branch. 
Analytical sensitivity functions are also derived for the dependence of these bounds on the degree of coupling between the base and the branch. 
The sensitivity analysis reveals further novel findings including the role of the frequency derivative of the branch dynamic stiffness in providing a direct relation between the band-gap edge locations and variation in the coupling parameters$-$the mass and stiffness ratios between the base chain and the branch root.~In additional analysis, sub-Bragg bounds of an exact model comprising a one-dimensional continuous base$-$modelled as a rod$-$and a discrete branch are derived and shown to be tighter than those of the all-discrete model. 
Finally, the applicability of the derived bounds and sensitivity functions are shown to be valid for a corresponding full two-dimensional finite-element model of a pillared waveguide admitting out-of-plane shear waves.

\end{abstract}

\end{frontmatter}

\newcommand\bigzero{\makebox(0,0){\text{\huge0}}}

\section{Introduction}

A noticeable property of locally resonant acoustic or elastic metamaterials is the possibility of exhibiting band gaps at relatively low frequencies; that is, frequencies for which the corresponding wavelength may be substantially larger than the characteristic sizes of the substructure units from which the metamaterial is formed~\cite{liu2000locally}. While spatial periodicity is not necessary in this class of artificial materials, it is often enforced in modelling and design to allow for a dispersion-curves characterization using Bloch's theorem~\cite{bloch1929quantenmechanik}, or, as commonly referred to in the structural dynamics community, periodic structures theory~\cite{mead1975wave}. A local-resonance band gap\footnote{A local-resonance band gap is also commonly referred to as a hybridization band gap.} appearing in the waveguide's band diagram is generated by a coupling between a local resonance (LR) associated with the substructure and one or more dispersion curves corresponding to the underlying host medium forming the base of the metamaterial. This type of band gap contrasts with a Bragg-scattering band gap which stems from destructive interferences of waves with wavelengths on the order of the unit cell size$-$a prime feature of phononic crystals~\cite{Sigalas_SSC_1993,Kushwaha_PRL_1993,vasseur2001experimental}.~Locally resonant metamaterials$-$admitting acoustic or elastic waves$-$may be realized in three-~\cite{liu2000locally,liu2002three,liu2005analytic}, two-~\cite{goffaux2003two,hsu2007lamb} or one-dimensional~\cite{fang2006ultrasonic,Khales_JVA_2013,khajehtourian2014dispersion} forms with embedded resonating subsystems.~The reader is referred to Refs.~\cite{deymier2013acoustic,hussein2014,MA_SA_2016,phani2017dynamics} for extensive reviews and detailed coverage of dynamic properties and applications of both elastic metamaterials and phononic crystals.\\
\indent A specific class of locally resonant elastic metamaterials, pertaining in particular to one-dimensional (1D) or two-dimensional (2D) wave propagation, is one where the local resonators are introduced as \it branching \rm substructures, for example, elastic rings around a rod~\cite{xiao2012longitudinal} or a beam~\cite{yu2006flexural,liu2012wave} or elastic pillars or stubs standing on an elastic plate~\cite{wu2008evidence,pennec2008low,Oudich_NJP_2010,Wu_IEEE_2011,Xiao_JPD_2012,bilal2013trampoline,Zhang_JAP_2013}.~A simple mass-in-mass lumped-parameter model for 1D wave propagation has emerged as a canonical model for branched acoustic or elastic metamaterials~\cite{huang2009negative}, and more complex models have appeared examining the effects of damping~\cite{Hussein_JSV_2013,Bacquet_arxiv_2018,Aladwani_JAM_2021} and nonlinearity~\cite{Lazarov_IJNM_2007,khajehtourian2014dispersion}.~Numerous applications have been proposed that utilize local resonances stemming from various types of branched substructures including, for example, sound isolation~\cite{assouar2016acoustic}, elastic waveguiding~\cite{pennec2008low,oudich2010propagation},  topological insulation~\cite{Chen_AIPA_2017,Yin_PRB_2018}, and nanoscale phonon manipulation for thermal conductivity reduction~\cite{Davis_PRL_2014,Hussein_AFM_2020}.~A recent review article surveys research on elastic metamaterials and metasurfaces exhibiting surface vibrational resonances~\cite{Jin_ROPP_2021}.\\
\indent The proposition of intrinsic local resonances at the material level, initiated by Liu et al.~\cite{liu2000locally}, inspired the investigation of a variety of elastic metamatieral configurations and the unique band-gap and dynamic effective properties that emerge. For example, in the context of branched media, numerous studies examined the effects of the geometric and material properties of the attached resonators~\cite{wu2008evidence,pennec2008low,khelif2010locally,badreddine2012enlargement}, or the host medium~\cite{bilal2013trampoline,Zhang_JAP_2013}, on the local-resonance band gaps$-$how low and wide they are.~The effect of material properties on the interaction of a low-frequency locally resonant band gap with a Bragg band gap has also been investigated, showing transition from one type to the other~\cite{liu2012wave} and in some cases coalescence into a much larger band gap~\cite{Xiao2011formation,khajehtourian2014dispersion}.~Furthermore, the influence of local resonances on the corresponding frequency-dependent effective mass/density and/or stiffness/elasticity has been studied in depth, with particular focus on demonstration of negative effective properties~\cite{liu2005analytic,huang2009negative,huang2009wave,huang2010study,chang2018wave}.~The rich design space afforded by the attachment of branching resonators to different types of waveguides continues to encourage investigation of a wide range of configurational possibilities including, for example, a phononic-crystal resonator~\cite{oudich2018rayleigh}, cantilevered beam-like resonator~\cite{Xiao_JPD-AP_2014} and multiple distinct resonators in the unit cell~\cite{Xiao_PLA_2012,Xiao_JSV_STL_2012}.~Although the focus is mostly on the dispersion characteristics of these systems, the response in a finite, truncated setting has also been examined~\cite{claeys2013potential,sugino2016mechanism,sugino2017general,al2017formation} including studies of the response of an isolated attached resonator~\cite{El-Khatib_JSV_2005,mace2014discussion} or the energetics of a single resonator~\cite{huang2009wave}.\\
\indent Given the broad appeal of waveguides with branching local resonators, analytical research has been pursued seeking a more in-depth elucidation of the fundamental properties of LR band gaps.~Mathematical studies emerged with varying levels of complexity in the treatment of the resonating branch, discretely modeled as a single~\cite{Xiao2011formation,xiao2013flexural,khajehtourian2014dispersion}, double~\cite{yu2006flexural,huang2010band}, or multiple~\cite{xiao2012longitudinal,miranda2019flexural}  degrees-of-freedom (DOF) substructure.~Xiao et al.~\cite{Xiao2011formation} studied the problem of a string with periodically attached resonators to elucidate band-gap formation mechanisms.~Analytical expressions for the formation and coupling of LR and Bragg band gaps were derived, providing physical insight into the underlying mechanisms for each band-gap type.~The band-gap attenuation profile was also examined and shown to vary as the local-resonance band-gap location relative to Bragg band gaps was changed.~This approach was later applied to a rod-based~\cite{xiao2012longitudinal} and a beam-based~\cite{xiao2013flexural} metamaterial, and also used to obtain approximate closed-form expressions for the edges of the two lowest band gaps in rods and shafts as well as design guidelines for sub-Bragg local-resonance band gaps~\cite{xiao2020closed}.~It was also utilized by Guo et al.~\cite{guo2020interplay} for studying the interplay between local-resonance and Bragg band gaps in acoustic waveguides with periodic Helmholtz resonators.~The same approach was subsequently extended to finite-strain thin rods with periodic resonators sheding light on the effects of nonlinearity on local-resonance band gap formation~\cite{khajehtourian2014dispersion}. \\
\indent It is common in discrete models to feature a \it mono-coupled \rm resonating branch, i.e., a branch coupled to the base medium through only a single DOF~\cite{mead1975wave}.~As mentioned above, this branch may be modeled as a single- or multi-DOFs resonator whose influence may be determined by individually varying its masses and spring constants and finding the effects of these changes on the location and size of the local-resonance band gaps \cite{Xiao2011formation,xiao2012longitudinal,xiao2013flexural}.~With this approach, however, it is difficult to draw general conclusions pertaining to how the properties of the branch and its level of coupling to the base medium correlates with the characteristics of the local-resonance band gaps, let alone to determine the level of sensitivity of this correlation.\footnote{The sensitivity of the correlation to the degree of coupling between the branch and base medium has been observed to exhibit complex trends even for single-DOF resonators, and was seen to be nonlinearly dependent on the level of coupling itself, as well as the resonating branch parameters~\cite{wang2006quasi,huang2010study}.}~Few studies aimed to express the resonating branch influence on band-gap characteristics in terms of its effective dynamic properties$-$a more general approach.~Using modal analysis, a method has been presented for multiple local-resonance band gaps control using the branch poles (resonances) and zeros (antiresonances)$-$described by the structure's transfer function~\cite{sugino2017general}.~In another approach, Raghavan and Phani~\cite{raghavan2013local} used a receptance-based method to find band-gap edges without performing a Bloch unit cell analysis.~This approach is based on the notion that band-gap edges occur at the unit-cell natural frequencies under fixed- and free-end boundary conditions~\cite{mead1975wave,Xiao2011formation}.~Conclusions were made with respect to single-DOF branch parameters, but it was suggested that the approach can be extended to multi-DOF branches as well.~Alternatively, from a wave propagation perspective, the poles and zeros of the branch may be considered given that they induce enhanced or reduced wave transmission~\cite{djafari2001surface, el2008transmission,williams2015theory,oudich2018rayleigh}.~Analytical wave transmission analysis based on effective dynamic forms of the branching structure may be found in Refs.~\cite{mei2005wave,williams2015theory,rughunanan2020behaviour}.~For example, Williams at al.~\cite{williams2015theory} treated symmetric Lamb wave transmission in a pillared plate using an impedance term to describe the role of the individual pillar$-$and the local-resonance band-gap edges were found to lie near the pillar poles and zeros.~Rughunanan et al.~\cite{rughunanan2020behaviour}, on their part, investigated 1D mono-coupled waveguides with conservative discontinuities featuring mass-spring resonators and related the wave transmission to dispersion properties such as the propagation constant, stop-band width, etc.~Their formulation utilized a dynamic-stiffness term to describe the effect of the discontinuities and enabled them to examine how the bandwidth (and maximum bandwidth points) of a stop band varies with the dynamic stiffness values.

\indent In this paper, we present a mathematical formulation that establishes a rigorous analytical connection between the properties of a general multi-DOF branch, the branch-base coupling properties, and the local-resonance band gap characteristics $-$ including bounds on their edges and their sensitivity to the coupling properties.~We model the branch dynamics through an effective dynamic stiffness, but unlike prior approaches we do not use transmission analysis.~The relationship we develop between the branch effective dynamic properties and the characteristics of the local-resonance band gaps provides us with a fundamental understanding that readily extends to any arbitrary resonating branch even when modeled as a continuum. Furthermore, we complement our theoretical treatment with illustrations that provide a graphical connection between the dispersion band structure and the dynamical properties of the resonating branch.~The branch effective dynamic properties incorporates features such as the poles and zeros, and inherently accounts for any design changes in the mass and spring constants for the case of a discrete branch or the geometric and material properties for the case of a continuous branch.~Dynamic condensation of the branch by an effective dynamic stiffness is an intuitive strategy because this effective quantity may simply be considered as a spring with a frequency-dependent stiffness.~The effective dynamic stiffness may also be readily measured from experiments or finite-element (FE) analysis~\cite{meirovitch2010fundamentals}.~This approach offers physical insights unobscured by the multiple design parameters that represent a resonating branch.~It provides a significant reduction to the design-parameter space, which is particularly useful for elucidating band-gap sensitivities to branch-base medium coupling.\\ 
\indent The paper is organized as follows. 
Section~\ref{sec:model} provides dispersion analysis of a discrete branched unit cell using Bloch's theorem and our branch-substructuring approach.~Subsection~\ref{subsec:bounds} identifies fundamental bounds for the local-resonance band gaps.~The band-gap edges' sensitivity to the degree of coupling is then derived in~\autoref{subsec:sensitivity}.~Guidelines for tuning local-resoannce band-gap width, attenuation strength, and effective properties are given in \autoref{subsec:tuning}.~Section \ref{sec:ContBase} extends the study of local-resonance band gaps to branched metamaterial unit cells with a continuum base and establishes tighter bounds for the band-gap edges.~The applicability of the derived bounds to a 2D all-continuum model for a branched unit cell is discussed in~\autoref{sec:2D}.~Finally, concluding remarks are given in~\autoref{sec:conclusion}. 
%


\section{Wave dispersion of branched chain}
\label{sec:model}

\begin{figure}[htbp!]
\centering
\centering
        \includegraphics[scale=1]{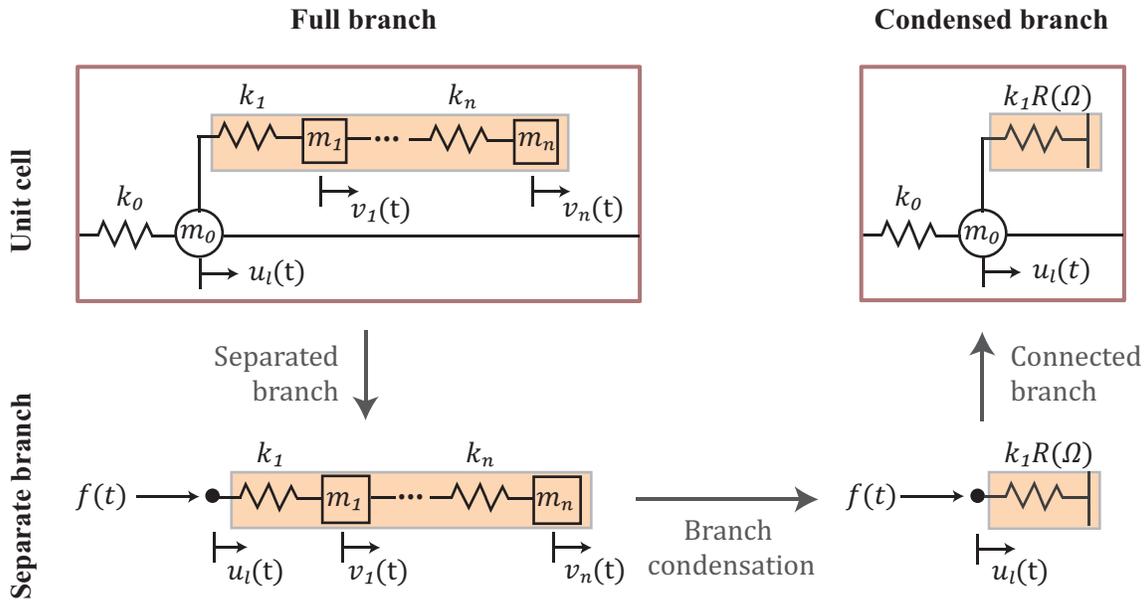}
    \caption{\textbf{(Top left)} A branched metamaterial unit cell with a base mass $m_0$ connected to an $n$-degrees-of-freedom resonating branch, which is inspected separately below \textbf{(bottom left)} with a point force applied at the previous connection node to the base. \textbf{(Bottom right)} The branch is condensed into an effective stiffness $R(\Omega)$ experienced by the point force and is replaced into the unit cell \textbf{(top right).}}
    \label{fig:Fig_01}
\end{figure}

Consider the branched metamaterial unit cell shown in the top-left corner of  \autoref{fig:Fig_01}. 
It contains a base chain with a lumped point mass $m_0$ moving with a displacement $u_l(t)$, where the subscript indicates the point mass index. 
Since this is an infinitely repeating unit cell, the point mass $m_0$ is connected to other equal masses (not shown in the figure), whose displacements are $u_{l-1}(t)$ and $u_{l+1}(t)$. 
The base-chain masses are inter-connected by springs with a stiffness constant $k_0$.
In the unit cell, the base-chain point mass is also connected to an $n$-DOF branch comprising lumped masses $m_j$ with displacements $v_j(t)$ and interconnected by springs of stiffness constants $k_j$,  where $j = 1, ..., n$.
In this section, the dispersion relation of the unit cell is derived. 
The resonating branch is condensed into its effective dynamic stiffness using the approach illustrated by the arrows in \autoref{fig:Fig_01} and explained in the upcoming subsections.


\subsection{Unit-cell dispersion analysis}
Starting at the top-left corner of \autoref{fig:Fig_01}, the point mass $m_0$ equation of motion can be written as  
\begin{equation}
m_{0} \ddot{u}_{l}=-k_{1}\left(u_{l}-v_1\right)-k_{0}\left(u_{l}-u_{l-1}\right)-k_{0}\left(u_{l}-u_{l+1}\right).
\label{eq:EOM1}
\end{equation}
We assume a traveling wave solution $u_l(t) = U_l e^{i (\kappa l a - \omega t)}$ and impose Bloch's boundary conditions $u_{l-1}(t) = u_{l}(t)e^{-i \kappa.a}$ and $u_{l+1}(t) = u_{l}(t)e^{i \kappa.a}$, where $\kappa$ is the wave number, $\omega$ is the temporal frequency and $a$ is the lattice constant. 
We also assume a harmonic wave solution for the branch masses $v_l(t) = V_l e^{i \omega t}$. Therefore, Eq. \eqref{eq:EOM1} can be rewritten as 
\begin{equation}
- m_{0} \omega^2 {U}_{l}= -k_{1}\left(U_{l}-V_1\right) - k_{0} U_{l} \left(1-e^{-i \kappa.a}\right)-k_{0} U_{l}\left(1-e^{i \kappa.a}\right).
\label{eq:EOM3}
\end{equation}
For convenience, we set $\Omega = {\omega}/{\Omega_1}$, where $\Omega_i = \sqrt{{k_i}/{m_i}}$, $\sigma = {k_0}/{k_1}$ and $\mu = {m_0}/{m_1}$. Then Eq. \eqref{eq:EOM3} can be expressed as 
\begin{equation}
- \mu \Omega^2 U_{l}=-(U_l - V_1) - \sigma U_{l} \left(1-e^{-i \kappa.a}\right)- \sigma U_{l}\left(1-e^{i \kappa.a}\right).
\label{eq:EOM4}
\end{equation}
It should be noted that the mass and stiffness ratios $\mu$ and $\sigma$ are inversely proportional to the degree of coupling between the base and the resonating branch.
This is because an increase in $\mu$ and $\sigma$ means that the base becomes heavier or stiffer relative to the branch, and hence the branch effect diminishes leading to a decrease in the degree of coupling. 

Finally, the system of linear equations describing the unit cell can be written as 
\begin{eqnarray}
\left[
\left(
    \begin{array}{cccccc}
    k_{f,1}+\sigma \left(2 - e^{i \kappa a} - e^{-i \kappa a} \right) & -k_{f,1}  & & & \bigzero \\                             
     -k_{f,1}  & k_{f,1}+k_{f,2}             & -k_{f,2}  & & \\
      &   -k_{f,2}            & k_{f,2}+k{f_3}  &  -k_{f_3} & \\
\\
     & \hspace{1cm}\ddots & \hspace{1cm} \ddots & & \hspace{-1cm}  \ddots   \\
     \\
  \text{\huge0}  & &  & -k_{f,n} & k_{f,n}
    \end{array}
 \right)
-
\Omega^2
\left(
    \begin{array}{cccccc}
 \mu & & & & \bigzero\\
 &  m_{f,1} & & & \\
  & & m_{f,2} & & \\
  \\
 \bigzero &  & & \ddots  &\\
 \\
 &  & & & m_{f,n}
    \end{array}
 \right)\right]
\begin{bmatrix}
U_l \\ V_1 \\V_2 \\ .\\ . \\ .\\ V_n 
\end{bmatrix} = 0,
\label{eq:ODEsys}
\end{eqnarray}
where $m_{f,j} = {m_j}/{m_1}$ and $k_{f,j} = {k_j}/{k_1}$. 

Substituting $q = \kappa a$, where $q$ is the normalized wave number, the eigenvalue problem in Eq. \eqref{eq:ODEsys} is solved for different values of $q$ spanning the first irreducible Brillouin zone (IBZ) to obtain the dispersion relation. 
A numerical example is presented in \autoref{fig:Fig_02} (top).~The real part of the dimensionless wave number $q$ indicates propagating waves, while the imaginary part indicates the exponential spatial attenuation rate. 
While the dispersion relation provides key information on the wave propagation characteristics of the model, in its raw form it does not reveal much of the underlying physical mechanisms from the perspective of the influence of the resonating branch.~In the next subsection, the branch is condensed into its effective dynamic stiffness to offer mechanistic insights into the role it plays in shaping the dispersion diagram and the LR band gaps it exhibits.


\subsection{Resonating branch effective dynamic stiffness}
\begin{figure}[htbp!]
\centering
\centering
        \includegraphics[scale=1]{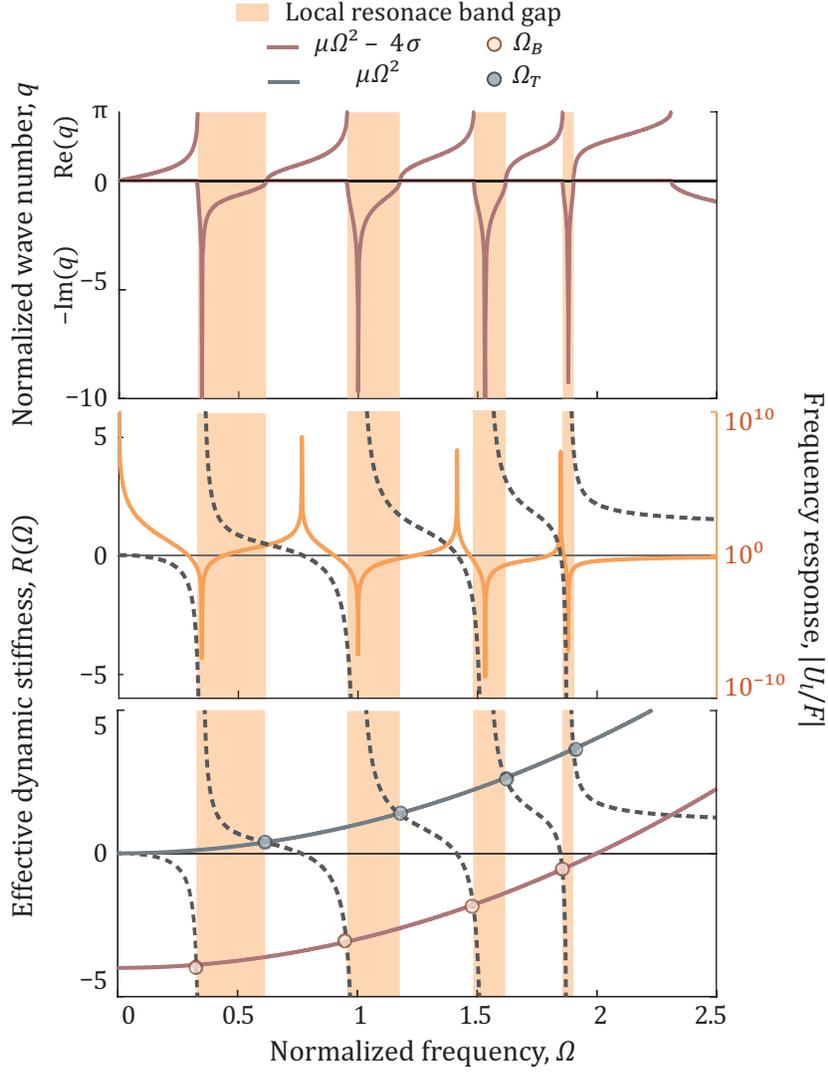}
    \caption{A numerical example for the branched metamaterial unit cell in \autoref{fig:Fig_01}, in which $\Omega_0 = 1$ rad/s, $\mu = 1, \, \sigma = 1$ and a 4-degrees-of-freedom resonating branch with mass ratios $m_{f,j} = 1$ and stiffness ratios $k_{f,j} = 1$, with $j = 1, ..., 4$, is connected to the base chain. \textbf{(Top)} Dispersion curves, \textbf{(middle)} left $y$-axis: effective dynamic stiffness $R(\Omega)$ and right $y$-axis: driven-point frequency response of the separate branch shown in the lower row of \autoref{fig:Fig_01}, \textbf{(bottom)} bottom  and top band-gap edges at the intersection points satisfying Eqs. \eqref{eq:OmBot} and \eqref{eq:OmTop}. }
    \label{fig:Fig_02}
\end{figure}

To express the resonating branch in terms of its effective dynamic stiffness, consider the separated branch in the lower left corner of \autoref{fig:Fig_01} with a force $f(t)$ acting on a massless node (indicated by the black circle) which was previously connected to the unit-cell base mass $m_0$.
Assuming a harmonic force $f(t) = F e^{i \omega t}$ and considering the steady state solution, we get  
 \begin{equation}
F - k_1 (U_l- V_1) = 0.
\label{eq:BranchEOM1}
\end{equation}
Using static condensation, $V_1$ can be expressed as a function of $U_l$,
\begin{equation}
V_1 = \zeta_{1}(\Omega) U_l,
\label{eq:V1rel}
\end{equation}
where $\zeta_{1}(\Omega)$ is obtained from the recursive formula
\begin{gather}
\zeta_{j}(\Omega)=\frac{\frac{k_{f,j}}{m_{f,j}}}{\left(\frac{k_{f,j}}{m_{f,j}}-\Omega^{2}\right)+\frac{k_{f,j+1}}{m_{f,j}}\left(1-\zeta_{j+1}(\Omega)\right)},
\label{eq:Condens}
\end{gather}
where the index $j = n, n-1,...,1$, starts at the branch last mass. Thus, $\zeta_{1}(\Omega)$ contains all the resonating branch parameters.
Note that in Eq. \eqref{eq:Condens}, $k_{f,n+1} = 0$ if the branch end is free as shown in \autoref{fig:Fig_01}, or $k_{f,n+1} \neq 0$ if it is fixed. Also, $\zeta_{n+1}(\Omega)$ is set to equal zero.
Inserting Eqs. \eqref{eq:Condens} and \eqref{eq:V1rel} into Eq. \eqref{eq:BranchEOM1} yields
\begin{gather}
F = k_1 R(\Omega) U_l, 
\end{gather}
where
\begin{gather} 
R(\Omega)=1-\zeta_1(\Omega).
\label{eq:EffectStiff}
\end{gather}
It can be seen that $R(\Omega)$ is the stiffness experienced by the force acting on the connection node normalized by $k_1$, the connecting spring between the base and branch in the unit cell in \autoref{fig:Fig_01}.
Moreover, the driven frequency response function of the branch at the connection node is written as 
\begin{subequations}
\begin{gather}
\frac{U_{l}}{F}=\frac{1}{k_{1} R(\Omega)},\label{eq:Response}\\ 
\left|\frac{U_{l}}{F} \right| = \left|\frac{1}{k_{1} R(\Omega)}\right|, \, \, \,\quad  \phi = \angle \left( \frac{U_{l}}{F}\right) = \arg \left(\frac{1}{k_{1} R(\Omega)}\right). \label{eq:FRF}
\end{gather}
 \label{eq:ResponseFRF}
\end{subequations}
The frequency response magnitude is shown on the right axis in \autoref{fig:Fig_02} (middle), while the effective dynamic stiffness $R(\Omega)$ is plotted on the left axis.  
It is noted that the effective dynamic stiffness has a reciprocal relationship with the frequency response as per Eqs.~\eqref{eq:ResponseFRF}. In other words, the poles (resonances) of the frequency response function are the roots of the dynamic stiffness $R(\Omega)$; and the zeros (antiresonances) are the singularity points of $R(\Omega)$, as is also shown in \autoref{fig:Fig_02} (middle). Hence, $R(\Omega)$ can also be expressed as
\begin{gather}
R(\Omega) = \frac{\prod_{i = 1}^{n}\left(\Omega^2 - p_i^2\right)}{\prod_{i = 1}^{n}\left(\Omega^2 - z_i^2\right)},
\label{eq:EffectStiffpz}
\end{gather}
where $p_i$ is the $i^\text{th}$ pole and $z_i$ is the $i^\text{th}$ zero, and $p_1 \leq z_1 \leq p_{2} ... \leq p_n \leq z_n$.~\footnote{This result is arrived at using: (1) The fact that the driven-point frequency response poles are the eigenvalues of the free-free branch, while the zeros are the eigenvalues of the fixed-free branch \cite{wahl1999significance}. (2) The eigenvalue interlacing theorem states that if a principal submatrix is obtained by deleting a row and a column from a Hermitian matrix, then the submatrix eigenvalues interlace with those of the bigger matrix \cite{hwang2004cauchy}. (3) The fixed-free branch equations are obtained by deleting the row and column corresponding to the fixed degree of freedom from the free-free branch equations.}
Therefore, the branch is condensed into its effective dynamic stiffness $R(\Omega)$, obtained by studying it separately from the rest of the unit cell.


\subsection{Unit-cell dispersion with incorporation of resonating-branch effective dynamic stiffness}

Going back to the dispersion relation, substituting Eqs.~\eqref{eq:V1rel} and~\eqref{eq:EffectStiff} and $q = \kappa a$ into Eq.~\eqref{eq:EOM4} and using Euler's identity, the dispersion relation can finally be written as 
\begin{equation}
- \mu \Omega^{2} U_{l}= - \left(R(\Omega)+2 \sigma(1-\cos (q))\right) U_{l}.
\label{eq:Dispersion}
\end{equation}
Eq.~\eqref{eq:Dispersion} describes the dispersion curves in \autoref{fig:Fig_02} (top) with the resonating branch represented in terms of its effective dynamic stiffness $R(\Omega)$. 
This is schematically illustrated in the top-right corner of \autoref{fig:Fig_01}, in which the resonating branch is regarded as a spring with a frequency-dependent stiffness.
From Eq. \eqref{eq:Dispersion}, the LR band-gap edges can be found by setting $q = \pi$ for the bottom edge, and $q = 0$ for the top edge. 
Thus, the set of $n$ bottom LR band-gap edges can be expressed as 
\begin{equation}
\{\Omega_{L,i}: R(\Omega)  = \mu \Omega^2 -  4 \sigma\},
\label{eq:OmBot}
\end{equation}
and the top band-gap edges as 
\begin{equation}
\{\Omega_{U,i}: R(\Omega) = \mu \Omega^2\}.
\label{eq:OmTop}
\end{equation}
These band-gap edge conditions are graphically depicted in \autoref{fig:Fig_02} (bottom), which shows that the bottom edges occur at the intersection of the dynamic stiffness $R(\Omega)$ with $\mu \Omega^2 - 4 \sigma$; and the top ones  at the intersection of $R(\Omega)$ with $\mu \Omega^2$. 
From a Newtonian force equilibrium perspective, this indicates that the bottom edge occurs when there is a balance between the base inertia $\mu \Omega^2 U_l$ on one side, and the base restoring force $4 \sigma U_l$ and the resonating branch restoring force $R(\Omega) U_l$ on the other. Meanwhile, the top edge occurs when the base inertia is balanced by the branch restoring force. Note that there is no influence of the base restoring force on the top edge. 

In the next section, we utilize the derived dispersion relation in Eq.~\eqref{eq:Dispersion} to establish theoretical results regarding bounds for the LR band-gap edges, as well as the edges' sensitivity to the mass and stiffness ratios $\mu$ and $\sigma$ and to develop insights for tuning the band gaps.


\section{Local resonance band-gap edge bounds, sensitivity, and tuning}

\subsection{Bounds on local-resonance band-gap edges}
\label{subsec:bounds}
\begin{figure}[htbp!]
\centering
\centering
        \includegraphics[scale=1]{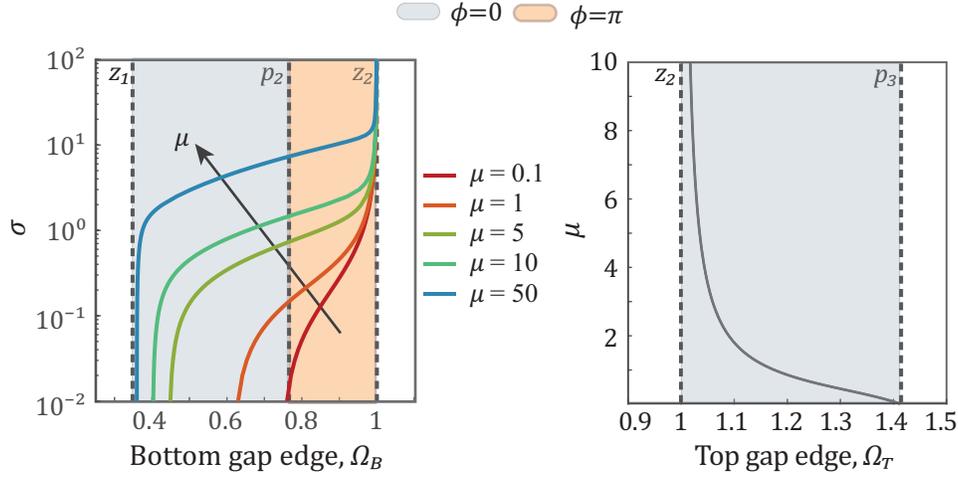}
        \caption{Variation of the bottom band-gap edge $\Omega_B$ \textbf{(left)} and top band-gap edge $\Omega_T$ \textbf{(right)} with mass and stiffness ratios $\mu$ and $\sigma$, as computed in a numerical parametric study for the second LR band gap. In all cases, a 4-DOF resonating branch with mass ratios $m_{f,j} = 1$ and stiffness ratios $k_{f,j} = 1$, with $j = 1, ..., 4$, is connected to the base chain.}
        \label{fig:Fig_03}
\end{figure} 

In this subsection, fundamental bounds on the branched metamaterial LR band-gap edges are derived in terms of the resonating branch dynamic characteristics.~For illustration, a numerical parametric study is performed in~\autoref{fig:Fig_03} by varying the mass and stiffness ratios $\mu$ and $\sigma$ and extracting the corresponding second LR band-gap edges.~Since we seek to understand the relationship between the gap edges and the resonating branch effective dynamic stiffness, it is useful to recall the definition of $R(\Omega)$ from Eq.~\eqref{eq:EffectStiffpz} from which we deduce that 
\begin{gather}
R(p_i) = 0, \quad
\lim_{\Omega\to z_{i}^-} R(\Omega) = -\infty, \quad \lim_{\Omega\to z_{i-1}^+} R(\Omega) = \infty. 
\label{eq:Rprop}
\end{gather}

\noindent \textbf{Local resonance band-gap bottom edge}
Looking at Eq.~\eqref{eq:OmBot}, it is seen that both sides can take either a positive or a negative sign. Therefore, the bottom  gap edge can fall into one of the following categories: 
\begin{itemize}
\item[1.] A case in which the base inertia $\mu \Omega^2$ is low or its restoring force $\sigma$ is high,  such that
\begin{equation}
\mu \Omega_{L,i}^2 - 4 \sigma < 0.
\label{eq:BelowBWcond}
\end{equation}

Then,  
\begin{gather}
R(p_i) > \mu \Omega_{L,i}^2 - 4 \sigma > \lim_{\Omega\to z_{i}^-} R(\Omega)  \quad
\Rightarrow \quad  p_{i} < \Omega_{L,i} < z_i.
\label{eq:BelowBW}
\end{gather}
Therefore, the LR band gap starts within an interval in which the phase angle $\phi = \pi$ [see Eq. \eqref{eq:FRF}], which means that the resonating-branch restoring force is in phase  with the base inertia force. This case is indicated by the orange region in \autoref{fig:Fig_03} (left). 

\item[2.] A case in which the base inertia $\mu \Omega^2$ is high or its restoring force $\sigma$ is low such that  
\begin{equation}
\mu \Omega_{L,i}^2 - 4 \sigma > 0. 
\label{eq:AboveBWcond}
\end{equation}
Then, using Eq. \eqref{eq:Rprop}, 
\begin{gather}
\lim_{\Omega\to z_{i-1}^+} R(\Omega) > \mu \Omega_{L,i}^2 - 4 \sigma > R(p_i),  \quad
\Rightarrow \quad  z_{i-1} < \Omega_{L,i} < p_i,
\label{eq:AboveBW}
\end{gather}
whereby the gap starts earlier than the first case, in a range of frequencies where the phase angle $\phi=0$. This means that the resonating branch restoring force is out of phase with the base inertia force. This frequency range is indicated by the grey area in \autoref{fig:Fig_03} (left).

\item[3.] The transition point between the first two cases when 
\begin{equation}
\mu \Omega_{L,i}^2 - 4 \sigma = 0 \quad \Rightarrow \quad \Omega_{L,i} = p_i.
\label{eq:AtBW}
\end{equation}
At this point, the resonating branch restoring force is null (see Eq. \eqref{eq:Rprop}) and the base inertia balances its own restoring force. This is also the point where the phase angle $\phi$ changes from $\pi$ to $0$. 
\end{itemize}

Interestingly, the transition point in case 3 also marks the bottom edge of a monatomic chain semi-infinite gap \cite{deymier2013acoustic,hussein2014}. 
This means that whether the LR band-gap bottom edge lies in the interval described by case 1 ($\phi = \pi$) or case 2 ($\phi=0$), depends on if the monatomic chain (without the resonating branch) experiences a pass band or a band gap at this frequency. 
For a more detailed contrast of the relationship between these cases and the monatomic lattice dispersion, see \ref{sec:appa}.

Combining these cases, it is concluded that for any positive-valued $\sigma$ and $\mu$, the $i^{\rm th}$ band gap caused by the resonating branch can start anywhere in the open interval 
\begin{equation}
z_{i-1} < \Omega_{L,i} < z_i, \quad i = 2, ... , n.
\label{eq:BotBound}
\end{equation}
In the numerical parametric study in \autoref{fig:Fig_03} (left), this is indeed shown to be the case. 
The numerical results also show that the stiffness ratio $\sigma$ exhibits a nonlinear relationship with the bottom edge frequency $\Omega_{B}$, which also depends on the mass ratio $\mu$. 

The first LR band gap represents a special case because the bottom edge will be bounded from below by $0$ since there is no preceding zero $z_{i-1}$, hence
\begin{equation}
0 < \Omega_{L,1} < z_1.
\label{eq:firstgap}
\end{equation}
However, the phase $\phi$ would vary depending on the boundary conditions.~If rigid body motion is allowed in the separated-branch [the $n^{\rm th}$ mass is free as depicted in the \autoref{fig:Fig_01} separate-branch schematic and implied by the plot of!\autoref{fig:Fig_02} (middle)], then $\phi= \pi$. Otherwise, if the branch is fixed at that end, then $R(0)>0$ and $\phi$ can either be $0$ or $\pi$ when the first gap starts.\\

\noindent \textbf{Local resonance band-gap top edge}
Moving on to the top band-gap edge and considering Eqs.~\eqref{eq:OmTop} and~\eqref{eq:Rprop}, for any positive real-valued $\mu$, 
\begin{equation}
\mu \Omega_{U,i}^2 >0 \quad \Rightarrow \quad  R(z_i) > \mu \Omega_{U,i}^2  > R(p_{i+1})\quad \Rightarrow \quad   z_{i} < \Omega_{U,i} < p_{i+1}, \quad i = 1, ..., n-1.
\label{eq:TopBound}
\end{equation}
This points to a region where the phase angle $\phi = 0$, which means that the branch restoring force is out of phase with the base inertia force. 
The numerical parametric study in \autoref{fig:Fig_03} (right) shows the top edge variation with the mass ratio $\mu$ for the second LR band gap, as well as the delimiting bounds. 

Therefore, the poles and zeros of the resonating branch constitute fundamental bounds for the intervals in which LR band-gap edges exist. In the next subsection, the band-gap edges' shift due to variation in the mass and stiffness ratios $\mu$ and $\sigma$ is examined analytically by deriving their sensitivity relations. This informs us about what happens within the edges' bounded intervals of existence.

\subsection{Sensitivity of LR band-gap edges}
\label{subsec:sensitivity}

The expressions in Eqs.~\eqref{eq:OmBot} and~\eqref{eq:OmTop} can be differentiated with respect to the mass and stiffness ratios $\mu$ and $\sigma$ in order to understand their effects on the location of the LR band-gap edges.  
Note that there is also a dependence on the resonating branch parameters.  
In general, 
\begin{gather}
R = R(\Omega; m_{f,1}, ..., m_{f,n}, k_{f,1}, ..., k_{f,n}),\\
\Omega_{B,i} = \Omega_{B,i}(\mu, \sigma;  m_{f,1}, ..., m_{f,n}, k_{f,1}, ..., k_{f,n}),\\
\Omega_{T,i} = \Omega_{T,i}(\mu, \sigma;  m_{f,1}, ..., m_{f,n}, k_{f,1}, ..., k_{f,n}),
\end{gather}
where the dependence of the band-gap edges on the branch parameters comes originally through $R(\Omega)$. 
It can be concluded from Eqs. \eqref{eq:EffectStiffpz} and \eqref{eq:Rprop}, and also seen in \autoref{fig:Fig_02}, that 
\begin{equation}
R'(\Omega) \leq  0,
\label{eq:Rderiv}
\end{equation}
for all frequency values. To simplify the notation, the derivative of $R(\Omega)$ with respect to frequency is denoted as $R'$. 
Taking the derivative of Eq. \eqref{eq:OmBot} with respect to the mass and stiffness ratios $\mu$ and $\sigma$, it is shown that 
\begin{equation}
\frac{\partial \Omega_B}{\partial \mu} = \frac{\Omega_B^2}{R' - 2 \mu \Omega_B},
\label{eq:SensMuBot}
\end{equation}
\begin{equation}
\frac{\partial \Omega_B}{\partial \sigma} = \frac{-4}{R' - 2 \mu \Omega_B}.
\label{eq:SensSigBot}
\end{equation}
Similarly, taking the derivative of Eq. \eqref{eq:OmTop} with respect to the mass and stiffness ratios $\mu$ and $\sigma$ yields, 
\begin{equation}
\frac{\partial \Omega_T}{\partial \mu} = \frac{\Omega_T^2}{R' - 2 \mu \Omega_T},
\label{eq:SensMuTop}
\end{equation}
\begin{equation}
\frac{\partial \Omega_T}{\partial \sigma} = 0.
\label{eq:SensSigTop}
\end{equation}
Looking at Eqs. \eqref{eq:SensMuBot}, \eqref{eq:SensMuTop} and \eqref{eq:Rderiv}, it can be seen that the top and bottom edge $\mu$-derivatives are always negative. This is in agreement with the results in subsection \ref{subsec:bounds}, which state that the top and bottom band-gap edges are closer to their upper bounds with decreasing values of $\mu$.
Similarly, looking at the bottom edge $\sigma$-derivative in Eq.~\eqref{eq:SensSigBot} and using Eq.~\eqref{eq:Rderiv}, it is seen that it is always positive. 
This is also in agreement with the fact, shown earlier in subsection \ref{subsec:bounds}, that the bottom gap edge comes close to the upper bound at high enough values of $\sigma$. 
Both of these trends can also be inferred from the graphical representation of the intersection points marking band-gap edges in \autoref{fig:Fig_02} (bottom).

Moreover, the bottom-edge derivatives with respect to $\mu$ and $\sigma$ have identical denominators, while the numerator of the $\mu$-derivative has a squared frequency term as opposed to a constant term in the $\sigma$-derivative. 
This means that the bottom-edge sensitivity to a perturbation in $\sigma$ is higher than to a perturbation in $\mu$ at low frequencies, i.e., $\sigma$ is more important than $\mu$ in determining the bottom edge at low frequencies. 
This is reversed at higher frequencies as the bottom edge sensitivity to $\mu$-perturbations gains prominence. 
Interestingly, all derivatives with the exception of Eq. \eqref{eq:SensSigTop} show an inverse relationship with $\mu$. 
Physically, this means that the sensitivity to perturbations increases with mass coupling, which is inversely proportional to the mass ratio $\mu$. 
It also indicates that the bottom-edge sensitivity to $\sigma$-perturbations depends on the value of $\mu$ (see \autoref{fig:Fig_03}), but not vice versa. Thus, these relations provide a mathematical understanding for the nonlinear trends in the numerical parametric studies in \autoref{fig:Fig_03} and the complex trends reported in the literature. 

Lastly, the sensitivity of the band-gap edges depends on the effective dynamic stiffness frequency derivative $R'$. Particularly, the sensitivity will increase with decreasing $R'$.
This can also be seen in the numerical example in \autoref{fig:Fig_02} (bottom) as having a flatter $R(\Omega)$ maximizes the shift in the intersection frequency resulting from perturbing either $\mu$ or $\sigma$. 
This sums up the effect of the resonating branch on the band-gap edges' sensitivity to perturbations in the mass and stiffness ratios. It also offers a new perspective for design problems motivating a focus on the objective of flattening the branch effective dynamic stiffness to achieve wider LR band gaps.~Additionally, it provides insights into the nature of the derived bounds.~Apart from the upper bound on the top edge that results from the physical impossibility of $\mu \leq 0$, it is seen that the sensitivity equals zero at the singularity points of $R(\Omega)$, which means that the band-gap edges will never pass these bounds. 
The next subsection offers some guidelines for tuning band-gap width, attenuation profile, and effective properties in light of the bounds and sensitivity results. 

\subsection{LR band-gap tuning}
\label{subsec:tuning}

\begin{figure}[htbp!]
\centering
\centering
        \includegraphics[scale=0.95]{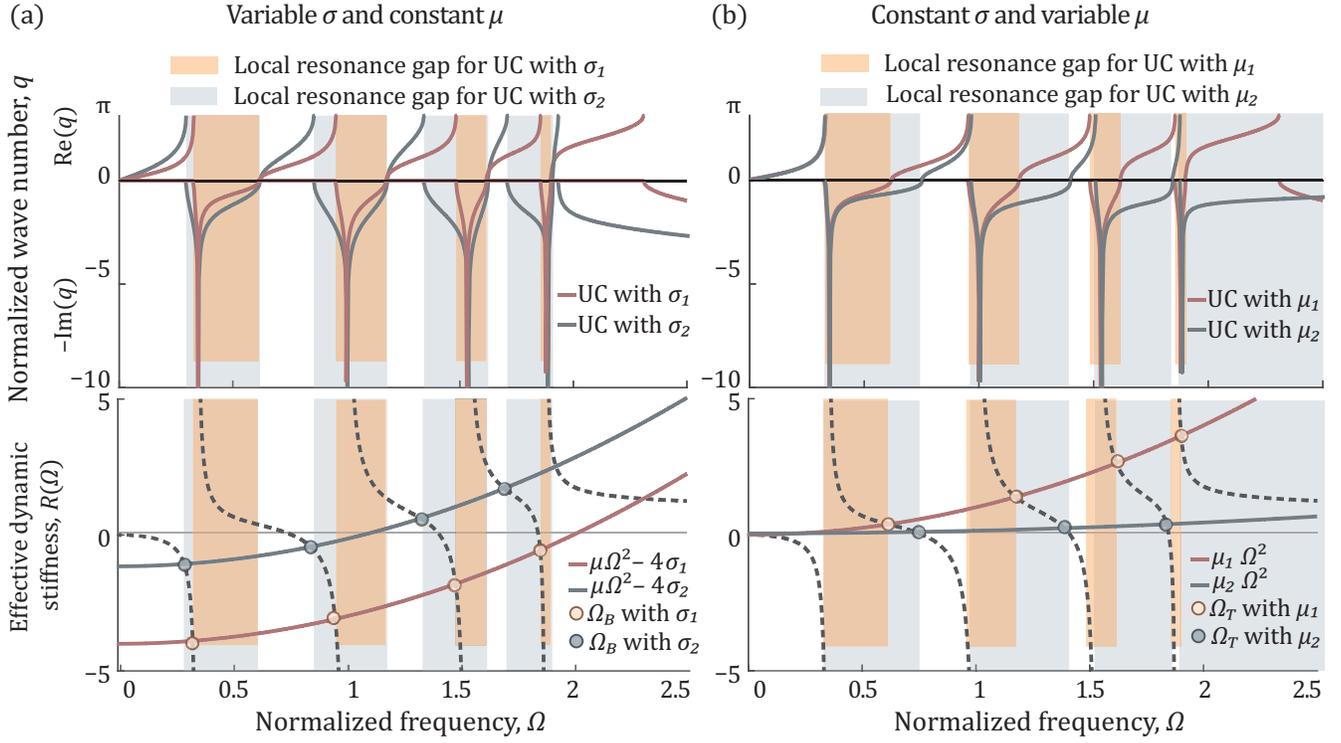}
        \caption{Two examples of the branched metamaterial unit cell in \autoref{fig:Fig_01} for a 4-DOF branch with $\Omega_1 = 1$, $m_{f,j} = 1$, $k_{f,j} = 1$ for $j = 1, ..., 4$.  \textbf{(a)} Dispersion curves (top) and intersection points (bottom) satisfying the bottom  band-gap edge condition in Eq. \eqref{eq:OmBot} with different stiffness ratios $\sigma_1 = 1$ and $\sigma_2 = 0.1$ and a fixed mass ratio $\mu = 1$. \textbf{(b)} Dispersion curves (top) and intersection points (bottom) satisfying the top band-gap edge condition in Eq.~\eqref{eq:OmTop} for different mass ratios $\mu_1 = 1$ and $\mu_2 = 0.1$ and a fixed stiffness ratio $\sigma = 1$.}
        \label{fig:Fig_04}
\end{figure}

\noindent \textbf{{Band-gap width}}\hspace{0.4mm} Following from the previous subsections, the band-gap width can be increased by locating the edges farther apart from each other within their allowed intervals.~In~\autoref{fig:Fig_04}(a), a numerical example shows the effect of decreasing the stiffness ratio $\sigma$ on the bottom edge. 
As predicted by the sensitivity relations in Eq. \eqref{eq:SensSigBot}, it is seen that the bottom edge is shifted downwards farther from the upper bound $z_i$ in the second case (grey) than in the first case (orange), thus widening the gap. This shift increases when the effective stiffness average slope is flatter in the neighborhood of the first-case bottom edge (orange).  
Another approach to widen the LR band gaps is presented in the numerical example in \autoref{fig:Fig_04}(b), in which the mass ratio $\mu$ is decreased instead.  
This causes the top band-gap edge to shift upwards closer to the upper bound $p_{i+1}$ in the second case (grey) than in the first case (orange), as predicted by Eq.~\eqref{eq:SensMuTop}. 
Also, note that this shift grows with the frequency which is also predicted by the quadratic frequency term in the numerator in Eq.~\eqref{eq:SensMuTop} (as opposed to the linear term in the denominator). The edge frequency shift also grows with flatter slopes of the effective stiffness $R(\Omega)$ in the neighborhood of the band-gap top edge (see the first gap for example).
On the other hand, decreasing $\mu$ in the second case (grey) has little effect on the bottom edge, which is already close to its upper bound $z_i$ and has a sensitivity close to zero [see Eq.~\eqref{eq:SensMuBot}]. 

It is worth noting that when considering any individual case in \autoref{fig:Fig_04}, the edges will move towards their lower bounds with higher band gaps. This is due to the growth of the base inertia term with frequency, compared to its constant stiffness term. This growth causes the edges to migrate towards their lower bounds [see Eqs.~\eqref{eq:TopBound} and~\eqref{eq:BotBound}]. This trend may seem contradictory to the earlier discussion concerning the sensitivity, in which the shift (caused by the $\mu$-decrease) of the top edges towards their upper bounds was magnified by increasing frequency.~However, that discussion is applicable to a specific gap that was being tuned by changing $\mu$, while the latter remark concerns band gaps with increasing order in the same case.~This apparent contradiction in the frequency role offers an instructive example that points to the complexity of predicting the band-gap edges' locations and trends without a guiding mathematical framework. \\

\noindent \textbf{{Spatial attenuation rate}}\hspace{0.4mm} Another objective of LR band-gap design is the strength of spatial attenuation, i.e., the imaginary component of the wave number $q$. 
It is known that the LR band-gap attenuation rate is maximum at the branch zero $z_i$ \cite{Xiao2011formation}.
This can be seen by rearranging Eq. \eqref{eq:Dispersion} as
\begin{equation}
 -\left(\mu \Omega^{2}-R(\Omega)-2 \sigma\right) U_{l}=  -2 \sigma \cosh (i q) U_{l}.
\label{eq:DispersionBG}
\end{equation}
Recalling that $q$ is imaginary inside the band gap and using Eq. ~\eqref{eq:Rprop}, then $q(z_i) \to \pm i \infty$, resulting in theoretically infinite attenuation at the resonating branch zero \cite{Xiao2011formation}.
Hence, the branch zero $z_i$ location within the band gap determines the attenuation rate profile, i.e., positioning it towards the middle or the extremities of the band gap results in a symmetric or an asymmetric attenuation profile, respectively. 
Recall that the branch zero $z_i$ is also the top-edge lower bound and the bottom-edge upper bound.

In the numerical example plotted in \autoref{fig:Fig_04}(a), the attenuation profile develops from the first until the last gap in the two cases due to the growing role of the inertia term $\mu \Omega^2$, which moves the edges towards their lower bounds, and thus shifts their location with respect to the maximum attenuation at $z_i$. 
In the first case (orange), the attenuation profile is mainly asymmetric in the first two band gaps and becomes more symmetric in the last two narrow band gaps.
In the second case (grey), lowering the stiffness ratio $\sigma$ shifts the band-gap bottom edge  downwards away from the maximum attenuation at the upper bound $z_i$. 
This causes the maximum attenuation to occur towards the band-gap middle in the second and third band gaps, thus resulting in a more symmetric attenuation profile. 
Due to the growing inertia role in the last band gap in the second case (grey), the bottom and top edges are closer to the lower bounds, $z_{i-1}$ and $z_i$, resulting again in an asymmetric attenuation profile with the maximum attenuation instead appearing closer to the top edge. 
Therefore, the knowledge of bounds and of the competing roles of the base inertia and stiffness are key to tuning the LR band-gap attenuation profile. \\

\noindent \textbf{{Effective dynamic properties}}\hspace{0.4mm} One last aspect considered in band-gap design is the effective negative properties and the mechanism of wave attenuation inside the band gap.
It is known that a negative effective dynamic mass occurs after the branch zero $z_i$ \cite{huang2009negative}. 
Therefore, to maximize regions of negative effective dynamic mass within the gap, it is desirable not only to widen the LR band gap, but lay most of it after $z_i$, such as in the second case (grey) in \autoref{fig:Fig_04}(b). 
For further discussion on the variation of the negative effective dynamic mass and its relation to the attenuation mechanism inside the band gap, see \ref{sec:appb}. \\

In summary, general guidelines for the design of band gaps can be drawn from the knowledge of the band-gap edge bounds and sensitivity to perturbations in the branch-base medium coupling, as well as how the attenuation profile and effective dynamic mass vary relative to the branch zeros. 
This concludes the branched metamaterial unit-cell discrete model results and discussion.~These results are naturally applicable to physically discrete media, such as granular systems exhibiting local resonances~\cite{gantzounis2013granular,matlack2018designing}.~The next sections examine their applicability in the continuum regime, and hence their value for continuous media treated either analytically or by FE modeling.


\section{Comparison to the sub-Bragg bounds of an exact model comprising a 1D continuum base and a discrete branch}
\label{sec:ContBase}

\begin{figure}[htbp!]
\centering
\centering
        \includegraphics[scale=1]{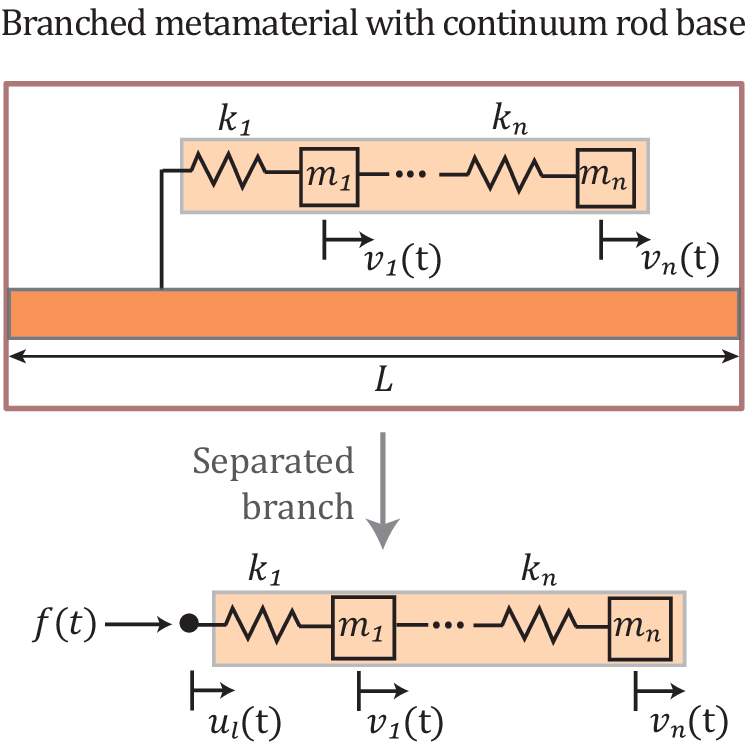}
        \caption{\textbf{(Top)} A branched metamaterial unit cell with a continuum base modeled as a 1D rod connected to an $n$-degrees-of-freedom resonating branch, which is inspected separately \textbf{(bottom)} with a point force applied at the previous connection node to the base.}
        \label{fig:Fig_05}
\end{figure}

In this section, we show that in a continuum-base/discrete-branch model of a branched metamaterial, in the sub-Bragg frequency domain, the lower bound for the $i^\textrm{th}$ LR band-gap bottom edge is the $i^\textrm{th}$ pole of the resonating branch, $p_i$, instead of $z_{i-1}$. 
This is justified by showing that the frequency corresponding to a resonating-branch pole actually satisfies the conditions for a Bragg-band-gap edge. 
Therefore, the sub-Bragg LR resonance bottom band-gap edge cannot lie on a resonating-branch pole; otherwise, the band gap is no more in the sub-Bragg regime and the wave-attenuation is due to both the local-resonance and Bragg effects. 

We consider a rod with periodic resonating branches and follow an approach that is based on the analytical treatment presented by Xiao et al.~in Ref.~\cite{Xiao2011formation} for a string with periodic resonators. 
Consider a rod unit cell of length $L$, Young's modulus $E$ and density $\rho$ with periodic resonating branches of $n$ DOFs, as shown in \autoref{fig:Fig_05}. Using the transfer matrix method \cite{mead1996wave,liu2012wave}, the rod unit-cell dispersion equation can be written as 
\begin{gather}
\cos(\kappa L) - \cos\left(\frac{\omega L}{c}\right) = \frac{k_1 R(\omega)}{2 \omega \rho c} \sin\left(\frac{\omega L}{c}\right)
\label{eq:AppBeq1}
\end{gather} 
where $c$ is the long-wave longitudinal speed and $k_1 R(\omega)$ is the resonating branch non-normalized effective dynamic stiffness (see Eq.~\eqref{eq:Response}). 

Setting $\kappa L = q$, where $q$ is the normalized dimensionless wave number and considering that the band gaps occur at the edge of the first IBZ, the band-gap edges' equation can be written as 
\begin{gather}
\cos\left(\frac{\omega}{c} L\right) + \frac{k_1 R(\omega)}{2 \omega \rho c} \sin\left(\frac{\omega}{c} L \right) = \pm 1.
\label{eq:AppBeq2}
\end{gather} 
Setting 
\begin{equation}
P = \frac{k_1 R(\omega)}{\omega \rho c},
\label{eq:AppBeq2b}
\end{equation}
and $\bar{\Omega} = {\omega L}/{\pi c}$, and squaring Eq. \eqref{eq:AppBeq2}, we obtain
\begin{gather}
\sin(\bar{\Omega} \pi)\left( P^2 \sin(\bar{\Omega} \pi) - 4 P \cos(\bar{\Omega} \pi) - 4 \sin(\bar{\Omega} \pi)  \right)=0.
\label{eq:AppBeq3}
\end{gather} 
Upon further manipulation using trigonometric identities, Eq. \eqref{eq:AppBeq3} can be rewritten as 
\begin{equation}
\begin{aligned}
4 \sin \frac{\bar{\Omega} \pi}{2} \cos \frac{\bar{\Omega} \pi}{2}\left(P \sin \frac{\bar{\Omega} \pi}{2}-2 \cos \frac{\bar{\Omega} \pi}{2}\right)\left(P \cos \frac{\bar{\Omega} \pi}{2}+2 \sin \frac{\bar{\Omega} \pi}{2}\right) &=0.
\end{aligned}
\label{eq:AppBeq4}
\end{equation}
Xiao et al. \cite{Xiao2011formation} have shown that the roots of Eq. \eqref{eq:AppBeq4} indicate the edges of band gaps caused by both LR and Bragg scattering effects.~The following relations
\begin{gather}
\sin \frac{\bar{\Omega} \pi}{2}=0, \quad \bar{\Omega} = \frac{\omega L}{\pi c} = n,\quad n = 2, 4, 6 ... 
\label{eq:AppBeq5}
\end{gather}
and 
\begin{gather}
\cos \frac{\bar{\Omega} \pi}{2}=0 ,\quad \, \bar{\Omega} = \frac{\omega L}{\pi c} = n,\quad n = 1, 3, 5 ... 
\label{eq:AppBeq6}
\end{gather}
indicate band-gap edge frequencies corresponding to integral multiples of half the wavelength between the resonating branches. This is entirely dependent on the base parameters and its length $L$, hence it is relevant to a Bragg-type mechanism. While the roots of  
\begin{equation}
P \sin \frac{\bar{\Omega} \pi}{2}-2 \cos \frac{\bar{\Omega} \pi}{2} = 0,
\label{eq:AppBeq7}
\end{equation}
and 
\begin{equation}
P \cos \frac{\bar{\Omega} \pi}{2}+2 \sin \frac{\bar{\Omega} \pi}{2}= 0,
\label{eq:AppBeq8}
\end{equation}
define band-gap edges caused by the resonating branch parameters in $P$. This includes low-frequency sub-Bragg LR band gaps. 

Now, let us approximate Eqs.~\eqref{eq:AppBeq7} and \eqref{eq:AppBeq8} in the low-frequency regime corresponding to $\bar{\Omega} = {\omega L}/{\pi c} \approx 0$ and substitute the value of $P$ from Eq. \eqref{eq:AppBeq2b}. After some manipulation and using the normalized frequency $\Omega = {\omega}/{\sqrt{k_1/m_1}}$, Eq. \eqref{eq:AppBeq7} becomes 
\begin{equation}
R(\Omega) = \frac{4 \rho c^2}{L} = 4 \sigma_{\textrm{cb}},
\label{eq:AppBeq7approx}
\end{equation}
where $E = \rho c^2$ and $\sigma_{\textrm{cb}} = {E}/{k_1 L}$ is the analog of the stiffness ratio $\sigma$ we previously used in the discrete model.~Note the similarity between Eq.~\eqref{eq:AppBeq7approx} and the discrete-model bottom-band-gap-edge equation in Eq. \eqref{eq:OmBot}. However, unlike Eq.~\eqref{eq:OmBot}, it does not have an inertia term. 
This can be attributed to the fact that at very low values of $\bar{\Omega} = {\omega L}/{\pi c} \approx 0$, both the rod density $\rho$ and the frequency $\omega$ can have low values, making the inertia term negligible. 
The same procedure can be performed with Eq. \eqref{eq:AppBeq8}, resulting in 
\begin{equation}
R(\Omega) =  \mu_{\textrm{cb}} \Omega^2,
\label{eq:AppBeq8approx}
\end{equation}
where $\mu_{\textrm{cb}} = {\rho L}/{m_1}$, which is the analog of the mass ratio $\mu$ previously used in the discrete model.~Equation~\eqref{eq:AppBeq8approx} matches the top-band-gap-edge equation Eq. \eqref{eq:OmTop} of the discrete model. 

\begin{figure}[htbp!]
\centering
\centering
        \includegraphics[scale=1]{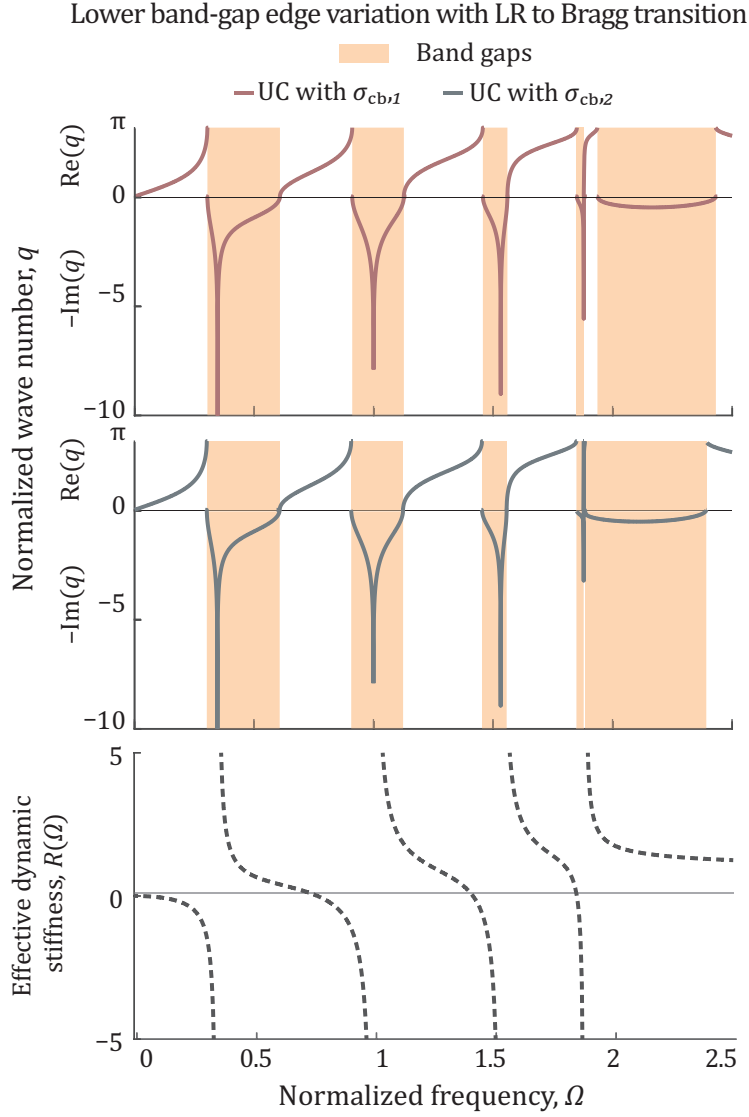}
        \caption{Two examples of the branched metamaterial unit cell in \autoref{fig:Fig_05} with $\mu_{\textrm{cb}} = 1$ and a 4-DOF branch with $\Omega_1 = 1, m_{f,j} = 1, k_{f,j} = 1$ for $j = 1,...,4$ are shown in the top and middle. \textbf{(Top)} Dispersion curves are plotted for the unit cell with stiffness ratio $\sigma_{\textrm{cb}} = 0.38$. \textbf{(Middle)} Dispersion curves are plotted for the unit cell with stiffness ratio $\sigma_{\textrm{cb}} = 0.36$; \textbf{(Bottom)} Effective dynamic stiffness $R(\Omega)$ of the resonating branch with the singularity points representing branch zeros and roots representing branch poles. The resonating-branch $4^{\textrm{th}}$ pole bounds the LR band gap from below as the Bragg band gap closes in with decreasing stiffness ratio, $\sigma_{\textrm{cb}}$. }
        \label{fig:Fig_06}
\end{figure}

Moving on to discuss bounds on band-gap edges in the continuum-base model, the same reasoning can be used as in the discrete-model case to conclude that the top-band-gap-edge bounds stay the same and can be written as 
\begin{equation}
z_i < \Omega_{T,i} < p_{i+1},
\label{eq:UppBoundContin1D}
\end{equation}
where $i$ stands for the band-gap index. 
However, different bounds can be obtained for the bottom band-gap edge as it is noted that $4 \sigma_{\textrm{cb}} > 0 $, resulting in 
\begin{equation}
p_i < \Omega_{B,i} < z_i.
\label{eq:LowBoundContin1D}
\end{equation}
This effectively tightens the bounds for LR band gaps in a branched metamaterial unit cell with a continuum base.
In fact, as we approach the lower bound by decreasing the value of $\sigma_{\textrm{cb}}$, thus decreasing the Young's modulus $E$, the assumption that $\bar{\Omega} = {\omega L}/{\pi c} \approx 0$ is no longer valid as the value of $c = \sqrt{{E}/{\rho}}$ decreases, causing $\bar{\Omega}$ to increase$-$this means that the frequency is approaching the Bragg-scattering regime. 
Moreover, it is recalled from Eq.~\eqref{eq:EffectStiffpz} that
\[ R(p_i) = 0,\quad i=1,..,n,\]
which leads to $P$ being $0$ [from Eq. \eqref{eq:AppBeq2b}].
Inserting this into Eqs. \eqref{eq:AppBeq7} and \eqref{eq:AppBeq8}$-$the exact expressions for band-gap edges affected by the resonator$-$the Bragg band-gap edges' conditions in Eqs. \eqref{eq:AppBeq5} and \eqref{eq:AppBeq6} are recovered. 
Therefore, it is concluded that the pole frequency $p_i$ of the resonating branch can coincide with a Bragg band gap.  
Therefore, the resulting band gap is no longer a pure sub-Bragg LR band gap, as is typical in the low-frequency regime. 
This is also depicted in \autoref{fig:Fig_06} as it shows the Bragg band gap approaching the $4^{\textrm{th}}$ LR band gap and finally merging with it as the value of $\sigma_{\textrm{cb}}$ decreases. 
The bottom band-gap edge never goes lower than the $4^{\textrm{th}}$ pole, which is the lower bound, unless the band gap changes from that of a LR to a mixed LR-Bragg band gap. 
Note that the transition of the band-gap attenuation mechanism from LR to mixed LR-Bragg as the resonator stiffness decreases was pointed out before by Liu et al. \cite{liu2002three}, Xiao et al. \cite{xiao2012longitudinal} and Liu and Hussein \cite{liu2012wave}.
However, establishing bounds for the band-gap edges before the onset of Bragg scattering effects as the stiffness decreases was not performed before, to the authors' knowledge.  
Needless to say, the sub-Bragg frequency regime is the operating range for most branched-metamaterial applications, making the model predictions useful for these applications. 

\section{Comparison to the bounds and sensitivity of a finite-element model comprising a 2D base and a 2D branch}
\label{sec:2D}

\begin{figure}[htbp!]
\centering
\centering
        \includegraphics[scale=1]{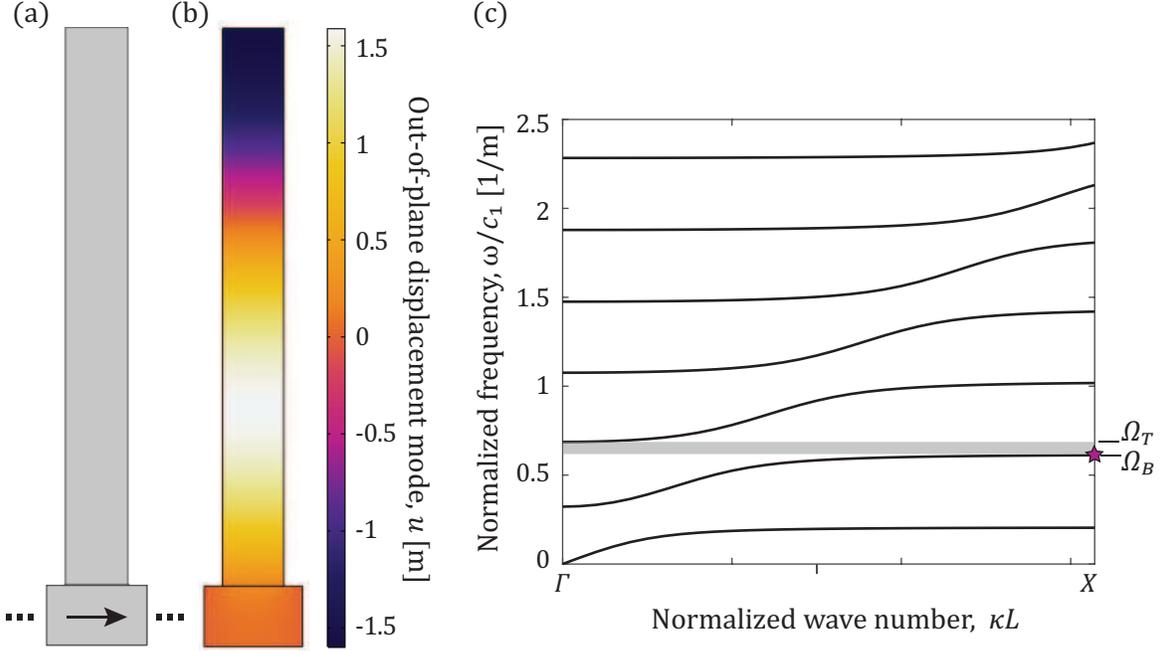}
        \caption{A branched/pillared continuum unit cell in which waves travel along the base as indicated by the arrow \textbf{(a)}, and corresponding dispersion curves plotted in \textbf{(c)} for a numerical example where the density ratio $\mu_{\text{cont}} = 2$, modulus ratio $\sigma_{\text{cont}} = 2.5$, branch Young's modulus $E_{\text{branch}} = 4 \, \text{GPa}$, branch density $\rho_{\text{branch}} = 1000 \, \text{kg/m}^3$, and base and branch Poisson ratio $\nu = 0.34$.~The mode shape at the second LR band gap bottom edge (marked by the star in the dispersion diagram) is shown in \textbf{(b)}. }
        \label{fig:Fig_07}
\end{figure}

In this section, we transition further to an all-continuum model where both the base and the branch are modeled as a continuum [see \autoref{fig:Fig_07}(a)].~This is often referred to as a pillared unit cell~\cite{Jin_ROPP_2021} in which waves propagate along the base and are influenced by the standing waves oscillating in the branching pillar.~We select a scalar continuum model that admits out-of-plane shear waves for which the governing partial differential equation is 
\begin{equation}
\nabla . \left(G(x,y) \nabla u(x,y,t) \right) = \rho(x,y) \ddot{u}(x,y,t),
\label{eq:Scalar}
\end{equation}
where $\rho(x,y)$ is the density, $G(x,y)$ is the shear modulus and $u(x,y,t)$ represents the displacement field. Constant material properties are assumed in each of the base and pillar domains. The density ratio $\mu_{\text{cont}} = {\rho_{\text{base}}}/{\rho_\text{branch}}$ and modulus ratio $\sigma_{\text{cont}} = {G_{\text{base}}}/{G_\text{branch}}$ are used to replace the mass and stiffness ratios $\mu$ and $\sigma$ in the discrete model. The Poisson ratio $\nu$ is assumed to be the same in the base and branch.  
An FE analysis is performed using COMSOL Multiphysics®, in which Bloch periodic boundary conditions are imposed to compute the unit-cell dispersion curves. A numerical example is plotted in \autoref{fig:Fig_07}(b) in which several LR band gaps are observed.
\begin{figure}[htbp!]
\centering
\centering
        \includegraphics[scale=1]{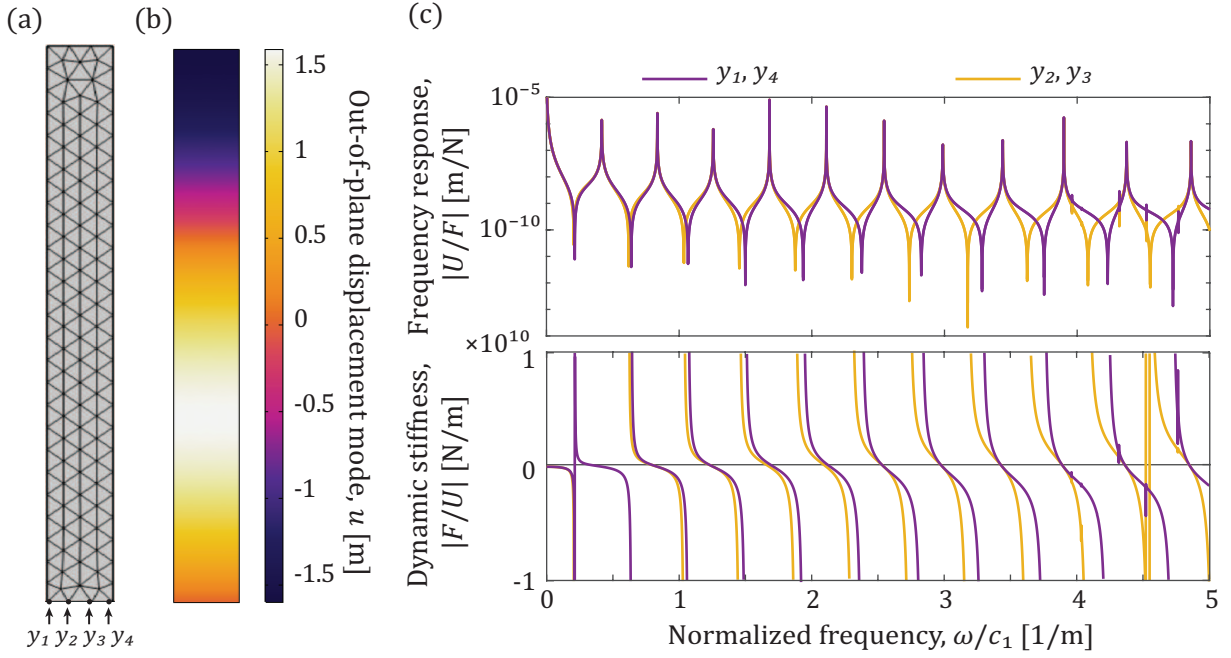}
        \caption{ \textbf{(a)} A meshed separate continuum branch showing nodal degrees of freedom $y_1, y_2, y_3$ and $y_4$ at the interface with the base where a distributed loading is applied. \textbf{(b)} Second fixed-free branch natural mode shape at $\omega/c =  0.6264$. \textbf{(c)} Branch frequency response (top) and effective dynamic stiffness (bottom) at different interface nodes for a numerical example where the Young's modulus $E_{\text{branch}} = 4 \, \text{GPa}$, density $\rho_{\text{branch}} = 1000 \, \text{kg/m}^3$ and Poisson ratio $\nu = 0.34$.}
        \label{fig:Fig_08}
\end{figure}

Following the same approach in \autoref{sec:model}, the resonating branch is separated as shown in \autoref{fig:Fig_08}(a).~To adapt the analysis to the 2D continuum domain, a distributed load is applied at the branch interface points $y_j$.~The frequency response and the effective dynamic stiffness at these points are plotted for a numerical example in \autoref{fig:Fig_08}(b) with the same values as in \autoref{fig:Fig_07}(b).~It can be seen that the first six LR band gaps shown in \autoref{fig:Fig_07}(b) take place around the first six branch zeros in \autoref{fig:Fig_08}(b). 
It is noted that the poles are aligned at the same frequency, but the zeros differ according to the measurement point $y_j$.~They grow further apart at higher frequencies as the wavelength becomes comparable to the interface points $y_j$ spacings.

On another note, a modal analysis is performed for the branch with fixed-free boundary conditions.~The separate branch second fixed-free mode shape occurs at a normalized frequency of $\omega/c = 0.6264$ m$^{-1}$, which lies close to the zero points in the frequency response plot in \autoref{fig:Fig_08}(b). This slight difference stems from the fact that the zeros are measured at the individual interface points, while the branch fixed-free mode shape takes place when all the interface points are fixed. Note the similarity between the separate branch fixed-free mode shape in the numerical example in~\autoref{fig:Fig_08} and the connected branch mode shape at the second band-gap bottom edge in \autoref{fig:Fig_07}.  

\begin{figure}[htbp!]
\centering
\centering
        \includegraphics[scale=1]{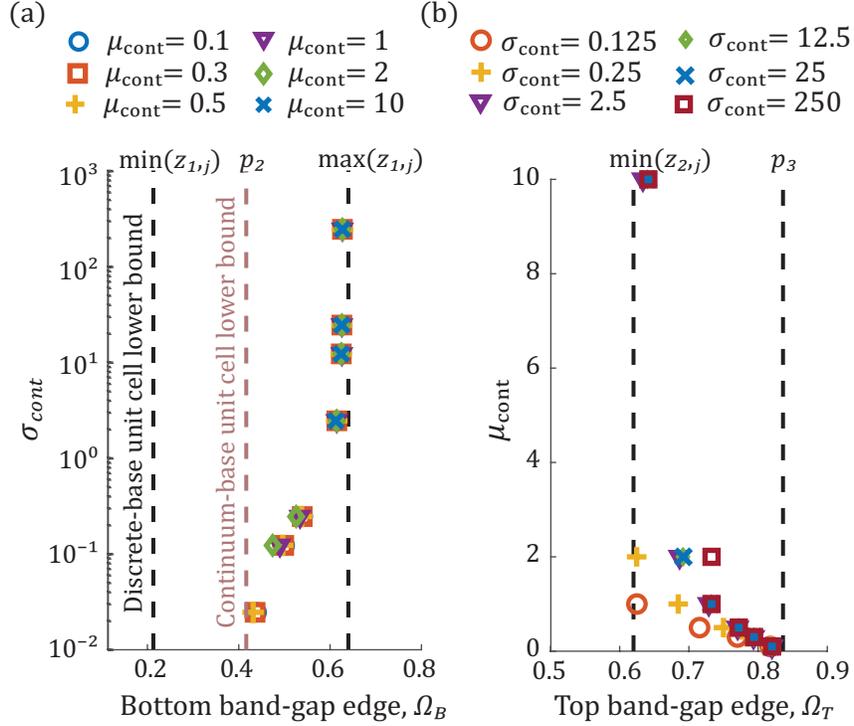}
        \caption{\textbf{(a)} FEM-calculated second band-gap bottom edges plotted against the modulus ratio $\sigma_{\rm cont}$ for different density ratios $\mu_{\rm cont}$. \textbf{(b)} FEM-calculated second band-gap top edges plotted against density ratio $\mu_{\rm cont}$ for different modulus ratios $\sigma_{\rm cont}$. In all cases, the branch Young's modulus $E_{\text{branch}} = 4 \, \text{GPa}$, density $\rho_{\text{branch}} = 1000 \, \text{kg/m}^3$ and Poisson ratio $\nu = 0.34$.}
        \label{fig:Fig_09}
\end{figure}

Finally, we perform a numerical parametric study in \autoref{fig:Fig_09} to investigate the effects of the density and modulus ratios $\mu_{\text{cont}}$ and $\sigma_{\text{cont}}$ on the band-gap edges.~Since the model predictions are generally expected to apply in the sub-Bragg regime, all data points for band gaps occurring after the onset of Bragg scattering effects are omitted.~Moreover, the bounds are modified to accommodate the multiple interface points by taking the minimum of all zeros for the lower bound and the maximum for the upper bound. Therefore, the bounds for sub-Bragg LR band-gap edges in a continuous branched metamaterial are expressed as  
\begin{equation}
p_i < \Omega_{B,i} < \max_j \, {z_{i,j}},
\label{eq:LowBoundContin}
\end{equation}
\begin{equation}
\min_j \, {z_{i,j}} < \Omega_{T,i} < p_{i+1},
\label{eq:UppBoundContin}
\end{equation}
where $i$ stands for the band gap index and $j$ stands for the measurement point index at the end of the branch. As can be seen in \autoref{fig:Fig_09} which focuses on the second LR band gap, the top and bottom LR band-gap edges are within the derived bounds. 
Moreover, the onset of the Bragg regime seems to occur whenever the bottom edge reaches the pole $p_2$ for different values of the density ratio $\mu_{\text{cont}}$ in \autoref{fig:Fig_09}(a). 
Thus, the $i^{\textrm{th}}$ pole constitutes the lower bound for the bottom band-gap edge of a pure LR band gap when the base is modeled as a continuous medium as predicted by Eq.~\eqref{eq:LowBoundContin1D}.~Moreover, the edges' locations variation with the density and modulus ratios is inversely proportional to the the effective dynamic stiffness slope in \autoref{fig:Fig_08} as predicted by the sensitivity relations in Eqs.~\eqref{eq:SensMuBot}, \eqref{eq:SensSigBot} and \eqref{eq:SensMuTop}. 
The bottom edge variation due to the modulus ratio $\sigma_{\rm cont}$ seems to be independent of the variation in the density ratio $\mu_{\text{\rm cont}}$ near the upper bound, as predicted by the continuum-base/discrete-branch model.~However, as the bottom edge approaches the lower bound, some dependence on the density ratio $\mu_{\text{\rm cont}}$ is noticeable, following the trend predicted by the all-discrete model.~Therefore, the all-discrete-model predictions better capture the inertia effects on the bottom-band-gap edge in the intermediate frequency range as Bragg scattering effects get closer.~The effect of the dynamic stiffness slope dominates the sensitivity of the band-gap edges to changing coupling parameters near the bounds as it approaches infinity.~On the other hand, the top edge $\Omega_T$ shows no dependence on the modulus ratio $\sigma_{\textrm{cont}}$ initially, as expected from the results of the all-discrete and continuum-base/discrete-branch models.~However, this gradually changes as the Bragg scattering regime gets closer, which is manifest in data points with $\sigma_{\text{\rm cont}}<1$.

With these results, we can confirm that the analytical predictions of the all-discrete and continuum-base/discrete-branch models$-$regarding the sensitivity of the band-gap edges and their bounding values$-$ hold for the all-continuum FE model in the sub-Bragg regime.~This affirms the applicability of our theoretical findings to LR band gaps in branched metamaterials in general.


\section{Conclusions}
\label{sec:conclusion}
\indent Considering 1D branched metamaterials, we have provided an analytical formulation$-$complemented by illustrative graphical representations$-$for relating the properties of a general multi-DOF branch, and the branch-base coupling parameters, to the LR band gap characteristics. Our formulation has yielded rigorous bounds on band-gap edges and their sensitivity to the coupling parameters. By examining the role of the resonating branch in terms of its effective dynamic stiffness, we obtain a more generalized understanding of its influence on LR band gaps, as opposed to investigating the direct influences of specific geometric and material properties. \\
\indent We started our analysis by focusing on an all-lumped parameter mass-spring unit-cell model representing a “typical” branched metamaterial.~Bloch analysis was performed to obtain the dispersion relation in terms of the resonating branch dynamic stiffness and the mass and stiffness ratios of the base to the branch first mass value and spring constant, respectively. Two sets of key physical findings are revealed: 

\begin{itemize}
  \item \it Bounds on band gap edges: \rm The top and bottom band-gap edges are proven to exist only within certain intervals, the bounds of which are shown to be described by the poles and zeros of the resonating branch.~For a given resonating branch configuration, the locations of band-gap edges within these intervals depend on the competition between the branch-to-base mass and stiffness coupling for a bottom band-gap edge, and solely on the mass coupling for a top band-gap edge.~Furthermore, the attenuation profile inside an LR band gap is determined by the band-gap edges' location relative to their respective bounds. For instance, if the bottom edge lies near the upper bound, or the top edge lies near the lower bound, it results in extreme asymmetric attenuation, and vice versa.
  \item \it Sensitivity to branch-base material coupling: \rm The derived sensitivity expressions for the locations of  band-gap edges with respect to perturbations in the degree of coupling show nonlinear dependence of the sensitivity on multiple parameters including the mass coupling and frequency. Among these parameters is also the frequency derivative of the resonating branch effective dynamic stiffness. The derivative dictates how sensitive the band-gap edge locations are to a given variation in the mass and stiffness ratios, with lower derivative values indicating higher sensitivity.  This points to a new criterion for designing optimal resonating branches given certain mass and stiffness ratios between the resonating branch and the base portion of the waveguide; such ratios may often be restricted by structural design requirements.\end{itemize}

\indent Lastly, it was analytically shown that the derived bounds hold for a sub-Bragg LR band gap when the base is modeled as a continuum. It was shown that a continuous base gives tighter bounds. The validity of our findings in the sub-Bragg frequency domain for a 2D FE-discretized all-continuum branched metamaterial model was also demonstrated by considering out-of-plane shear waves.

\section{Acknowledgments}

The authors gratefully acknowledge fruitful discussions with Professor Bahram Djafari-Rouhani. This research was partially supported by the Advanced Research Projects Agency-Energy (ARPA-E) grant number DE-AR0001056.


\appendix

\setcounter{figure}{0}  

\section{Relation between band-gap bottom edge phase $\phi(\Omega_{B,i})$ and monatomic lattice dispersion}
\label{sec:appa}

\begin{figure}[h!]
\centering
\centering
        \includegraphics[scale=1]{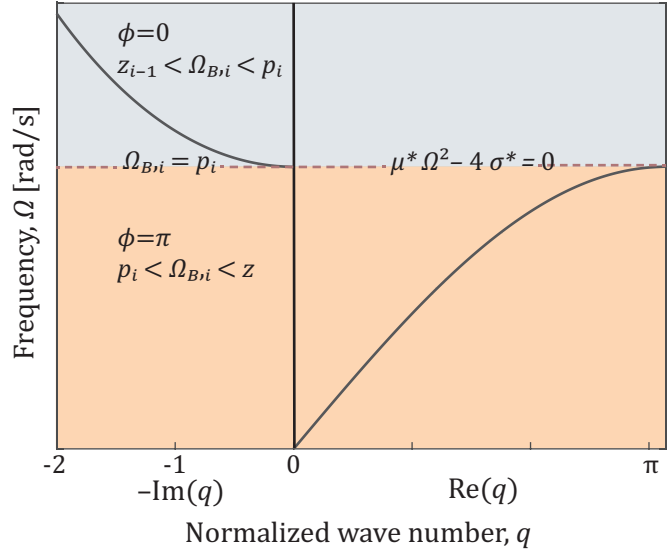}
        \caption{Dispersion of a monatomic lattice with point mass $\mu^*$ connected to a spring with constant $\sigma^*$. Shaded regions highlighting the bandwidth and semi-infinite band gap correspond to regions within the allowed intervals for branched metamaterial bottom band-gap edge as shown in \autoref{fig:Fig_03} and explained by Eqs. \eqref{eq:BelowBWcond}, \eqref{eq:AboveBWcond} and \eqref{eq:AtBW}.}
        \label{fig:Fig_A01}
\end{figure}
Wave propagation in a monatomic lattice is widely studied in the literature~\cite{deymier2013acoustic,hussein2014} and the dispersion equation can be expressed as 
\begin{equation}
\mu^* \Omega^2 U_l = 2\sigma^* (1-\cos(q)) U_l,
\label{eq:monatomic}
\end{equation}
for a monatomic lattice of point masses $\mu^*$ connected by spring constants $\sigma^*$.~Note the similarity of the inertia and stiffness terms between Eqs.~\eqref{eq:monatomic} and \eqref{eq:Dispersion}, with $\mu^*$ and $\sigma^*$ here representing the mass value and spring constant, as opposed to $\mu$ and $\sigma$ which represent the coupling parameters in Eq.~\eqref{eq:Dispersion}.~The dispersion relation is plotted in \autoref{fig:Fig_A01}. At the edge of the first IBZ, $q = \pi$ and a semi-infinite band gap starts at 
\begin{equation}
\mu^* \Omega^2 - 4 \sigma^* = 0.
\label{eq:monatomiclimit}
\end{equation} 
This condition coincides with Eq.~\eqref{eq:AtBW}, which corresponds to the bottom band-gap edge of a branched metamaterial falling on top of a branch pole.~Therefore, whether the branched metamaterial LR band-gap bottom edge starts within a region where the dynamic stiffness phase $\phi$ is $\pi$ or $0$ corresponds to whether the frequency falls within the monatomic lattice propagation band or band gap as can be seen in \autoref{fig:Fig_A01}. Note that the same color code is used for $\phi = \pi$ and $\phi=0$ regions as in \autoref{fig:Fig_03}.

\setcounter{figure}{0}

\section{Effective dynamic mass}
\label{sec:appb}
\begin{figure}[htbp!]
\centering
\centering
        \includegraphics[scale=0.95]{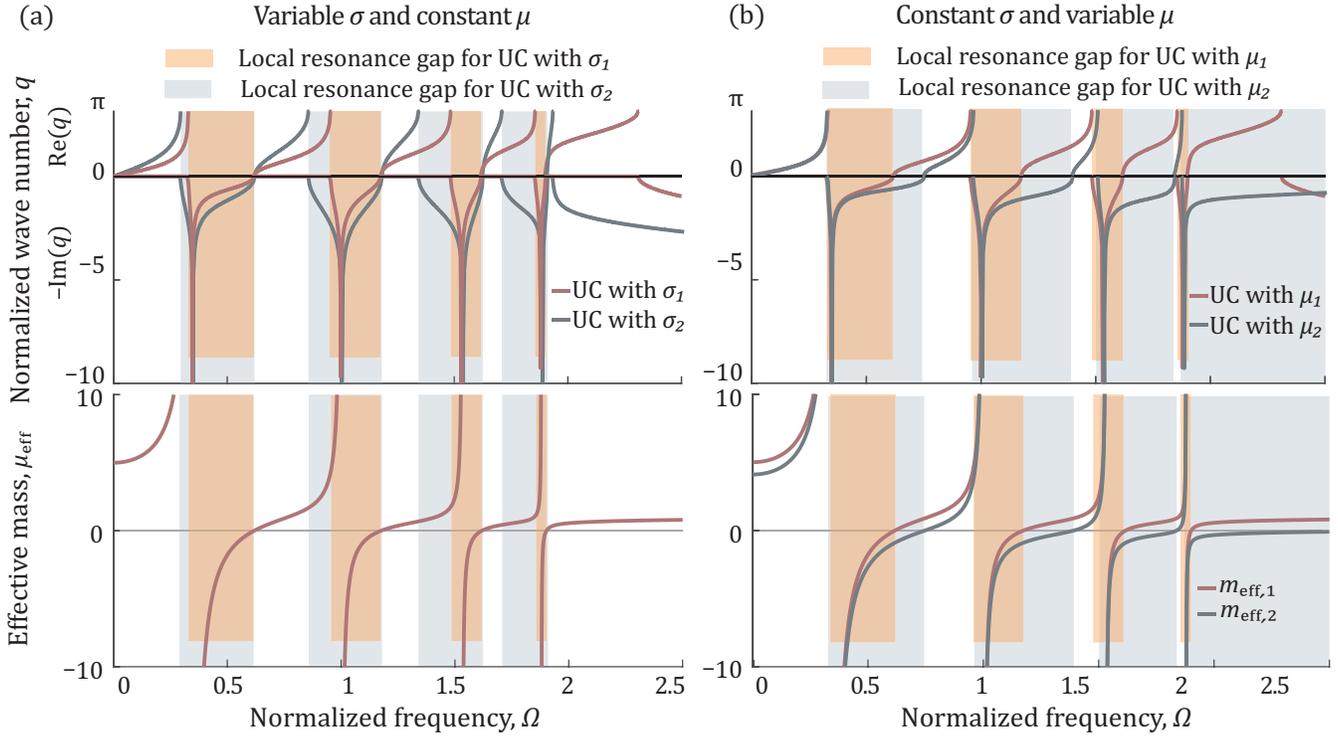}
        \caption{Two examples of the branched metamaterial unit cell in \autoref{fig:Fig_01} for a 4-DOF branch with $\Omega_1 = 1$, $m_{f,j} = 1$, $k_{f,j} = 1$ for $j = 1, ..., 4$.  \textbf{(a)} Dispersion curves (top) and effective dynamic mass (bottom) with different stiffness ratios $\sigma_1 = 1$ and $\sigma_2 = 0.1$ and a fixed mass ratio $\mu = 1$. Note that the effective dynamic mass is unaffected by the change in $\sigma$.  \textbf{(b)} Dispersion curves (top) and effective dynamic mass (bottom) for different mass ratios $\mu_1 = 1$ and $\mu_2 = 0.1$ and a fixed stiffness ratio $\sigma = 1$.}
        \label{fig:Fig_B01}
\end{figure}

From Eq.~\eqref{eq:Dispersion}, the effective dynamic mass ratio $\mu_{\text{eff}}$ can be written as 
\begin{equation}
\mu_{\text{eff}} = \mu - \frac{R(\Omega)}{\Omega^2}.
\end{equation}
Recalling Eq.~\eqref{eq:EffectStiffpz}, it is seen that the effective dynamic mass takes a negative sign after the branch zero $z_i$.~Therefore, from the discussion in \autoref{subsec:bounds}, it may be concluded that the effective dynamic mass can take positive or negative values inside the branched metamaterial band gap.~This is also in agreement with past work for single-DOF branches \cite{chang2018wave} and multi-DOF branches \cite{huang2010band}. 
Some numerical examples are shown in \autoref{fig:Fig_B01} for illustration.~It is seen that the band-gap regions before the branch zero $z_i$ exhibit a positive effective dynamic mass, while those shortly after $z_i$ exhibit a negative effective dynamic mass. 
Chang et al.~\cite{chang2018wave} have shown using numerical simulations that positive effective dynamic mass regions exhibit Bragg-like attenuation behavior while negative effective dynamic mass regions exhibit LR attenuation behavior, as previously discussed in the introduction section. They have determined that the Bragg-like attenuation behavior occurs if the LR frequency lies above the monatomic lattice pass band. It should further be noted that a positive effective dynamic mass can also occur when the branch zero $z_i$ is still within the monatomic lattice propagation band $\left(z_i <\sqrt{{4 \sigma^*}/{\mu^*}}\right)$ (from Eq.~\eqref{eq:monatomiclimit}) as observed in \autoref{fig:Fig_B01}.

  
\bibliographystyle{elsarticle-num-names}
\bibliography{litrev}

\end{document}